\documentclass{svmult}

\usepackage{makeidx}     
\usepackage{graphicx}    
\usepackage{multicol}    
\usepackage{url}
\usepackage{amsmath}
\usepackage{aasjournals}
\usepackage{natbib}
\bibpunct{[}{]}{;}{n}{}{,}

\makeindex             

\begin{document}

\title*{3D AMR Simulations of Point-Symmetric Nebulae}

\author{Erik-Jan Rijkhorst\and
Vincent Icke\and
Garrelt Mellema}
\institute{Sterrewacht Leiden, P.O. Box 9513, 2300 RA, Leiden, The Netherlands
\texttt{rijkhorst@strw.leidenuniv.nl}
}

%
\maketitle

\begin{abstract}
At the end of their lives low mass stars such as our Sun lose most of their mass.
The resulting planetary nebulae show a wide variety of shapes, from spherical to highly bipolar.
According to the generalized interacting stellar winds model, these shapes are due to an interaction between a very fast tenuous outflow, and a denser environment left over from an earlier slow phase of mass loss.
Previous analytical and numerical work shows that this mechanism can explain cylindrically symmetric nebulae very well.
However, many circumstellar nebulae have a multipolar or point-symmetric shape.
With two-dimensional calculations, Icke showed that these seemingly enigmatic forms can be easily reproduced by a two-wind model in which the confining disk is warped, as is expected to occur in irradiated disks.
Here, we present the extension to fully three-dimensional adaptive mesh refinement simulations of such an interaction.
\end{abstract}
%
%
\section{Introduction}
In the final phases of stellar evolution, low mass stars, such as our Sun, first swell up and shed a dense, cool wind in the asymptotic giant branch (AGB) phase.
This episode is followed by a fast, tenuous wind that is driven by the exposed stellar core, the future white dwarf.
The planetary nebulae (PNe) resulting from this expulsion phase come in a wide variety of shapes, from spherical to highly bipolar.
Some even have a multipolar or point-symmetric shape.
Balick \citep{1987AJ.....94..671B} proposed that such forms arise due to an interaction between a slow disk-shaped inner AGB nebula and the fast `last gasp' of the star.
Analytical \citep{1988A&A...202..177I, 1989AJ.....97..462I} and numerical \citep{1989ApJ...339..268S, 1991A&A...251..369I, 1991A&A...252..718M} work shows that this {\it generalized interacting stellar winds} (GISW) mechanism works very well (for an up-to-date review, see \citet{2002ARA&A..40..439B}).
Several scenarios for obtaining a disk around a PN exist, and it is in general assumed in the models that the shape of the dense gas around the star is a disk or a toroid.

Icke \citep{2003A&A...405L..11I} proposed that the point-symmetric shapes observed in a number of PNe are formed in an interaction between a spherical stellar wind and a {\it warped} disk.
It is prossible to produce such a warp around a single star, through the combined effects of irradiation and cooling \citep[e.g.][]{1996MNRAS.281..357P, 1996ApJ...472..582M}.
Whereas Icke's computations were restricted to a two-dimensional proof-of-principle, we now present a first series of fully three-dimensional hydrodynamic computations of such a wind-disk interaction using the technique of adaptive mesh refinement (AMR).
%
%
\section{Point-symmetric nebulae}
Work on cylindrically symmetric nebulae showed \citep{1988A&A...202..177I, 1991A&A...251..369I, 1992Natur.355..524I} that sharply collimated bipolar flows are a frequent and natural by-product of the GISW model.
However, many circumstellar nebulae have a multipolar or point-symmetric (i.e. antisymmetric) shape \citep{1993mlab.conf.....S, 1998AJ....116.1357S}.
The nebulae that are formed in the wind-disk interaction would naturally acquire the observed antisymmetry if the disk that confines the fast wind is warped, instead of symmetric under reflection about the equatorial plane.
In binary stars such warps might be common, due to various torques in the system, but few point-symmetric nebulae have been proven to have binary cores.
Several mechanisms have been proposed for warping a disk surrounding a {\it single\/} star.
The most interesting one for our purposes invokes radiative instability \citep{1977ApJ...214..550P, 1990A&A...239..221I, 1996MNRAS.281..357P, 1996ApJ...472..582M}.

Livio \& Pringle already proposed \citep{1996ApJ...465L..55L, 1997ApJ...486..835L} that the precession of warped disks might be responsible for point-symmetric nebulae.
They proved conclusively that the various physical scales for mass, accretion, luminosity and precession match the observations.
The production of the nebulae proper they attributed to an unspecified `jet' mechanism.
Observations of many bipolar nebulae with `ansae' (e.g. NGC~3242, NGC~7009) and `FLIERS' (e.g. NGC~6751, NGC~7662) leave little doubt that jets are occasionally formed during the evolution of some aspherical PNe, probably in the late post-AGB phase, before the fast wind has switched on.
But the nebulae presented by \citet{1998AJ....116.1357S} do not seem to resemble such shapes.
While leaving open the possibility that jets may be responsible for additional structures, as in the case of the `ansae', we show that the interaction between a warped disk and a spherically symmetric wind suffices.
The lobes of point-symmetric nebulae \citep{1993mlab.conf.....S, 1998AJ....116.1357S} look as if they were produced almost simultaneously.
This is difficult in the case of a precessing jet, which would make a corkscrew-like nebula of a type not readily apparent in post-AGB shells, although some objects do show features that are likely to be due to precession \citep{1993mlab.conf.....S}.
%
%
\section{Radiation driven warping}
When an accretion disk is subject to external torques it may become unstable to warping \citep{1975ApJ...195L..65B, 1977ApJ...214..550P, 1983MNRAS.202.1181P} and when irradiated by a sufficiently luminous central star even an initially flat disk will warp \citep{1990A&A...239..221I, 1996MNRAS.281..357P, 1996ApJ...472..582M, 1998ApJ...504...77M}.
The difference in radiation pressure on slightly tilted annuli at different radii will induce the warp.
Essential is that the disk is optically thick for the stellar radiation {\it and\/} for its own cooling flux.
The latter condition is the most restrictive, because a disk that is optically thin in the infrared will not suffice.
This restricts the disks to a specific subclass of high density and low temperature.

Analytical considerations lead to expressions for growth and precession rates and morphologies of the warp whereas numerical calculations including the effects of self-shadowing show that the non-linear evolution of the warp can produce highly distorted shapes, even with an inverted, counter-rotating inner disk region \citep{1997MNRAS.292..136P, 1999MNRAS.308..207W}.
Applications of warped disk theory range from active galactic nuclei \citep{1997MNRAS.292..136P, 1998ApJ...504...77M} to X-ray binaries \citep{1997ApJ...491L..43M, 1999MNRAS.308..207W}, protostellar disks \citep{1997ApJ...488L..47A}, and PNe \citep{1996ApJ...465L..55L, 1997ApJ...486..835L}.
Other mechanisms to produce warped disks, besides the radiatively driven one, are the wind driven \citep{2001ApJ...563..313Q} and the magnetically driven instability \citep{1999ApJ...524.1030L, 2003ApJ...591L.119L}.

Since we intend to study warped disks in PNe, we need a scenario to form accretion disks in these systems.
Plausible scenarios include the coalescence of compact binaries \citep{1994ApJ...421..219S}, absorption of planetary systems, or the formation of a disk due to a main-sequence companion being out of equilibrium when emerging from a common envelope (CE) phase with a primary AGB star.
In this case, the companion loses most of the mass it accreted during the CE phase which subsequently forms a disk around the primary \citep{1994ApJ...421..219S} and which, in a later stage, can get radiatively warped when illuminated by the central star of the PN.

For a PN the luminosity of the central star alone is sufficiently high to induce a radiation driven warp.
Following \citet{1997MNRAS.292..136P}, an expression for the radius $R_{crit}$ beyond which the disk is unstable to radiation driven warping is found from comparing the timescales of the viscous and radiation torques, leading to
\begin{equation}
\label{eq:critRadius}
  R_{crit} = (2\pi/A)^2 \; ,
\end{equation}
with the contant $A$ defined by $A^2 \equiv 1/4\,c^{-2} G^{-1} M_*^{-1} L_*^{2} \eta^{-2} \do
t{M}_{acc}^{-2}$.
Here $c$ is the speed of light, $G$ the gravitational constant, $\eta\equiv\nu_2/\nu_1$ is the ratio of the azimuthal to the radial viscosity, $M_*$ is the mass and $L_*$ the luminosity of the central star and $\dot{M}_{acc}$ is the disk's accretion rate.
We assumed a surface density $\Sigma_d=\dot{M}_{acc}/(3\pi\nu_1)$ \citep[e.g.][]{1981ARA&A..19..137P}.

In a cartesian coordinate system, the warped disk surface is given by \citep{1996MNRAS.281..357P}
\begin{equation}
\label{eq:diskSurface}
  \boldsymbol{x}(R,\phi)=
  R\left(
    \begin{array}{rcl}
       \cos\phi\sin\gamma & + & \sin\phi\cos\gamma\cos\beta\\
      -\cos\phi\cos\gamma & + & \sin\phi\sin\gamma\cos\beta\\
                          & - & \sin\phi\sin\beta
  \end{array}
  \right) \;,
\end{equation}
with local disk tilt angle $\beta(R,\phi)$, and orientation angle of the line of nodes $\gamma(R,\phi)$.
Here, $R$ and $\phi$ are the non-orthogonal radial and azimuthal coordinates respectively, pointing to the surface of the disk \citep[cf.][]{1996MNRAS.281..357P}.
In our model calculations we adopt the case of a steady precessing disk with no growth and zero torque at the origin for which we have in the precessing frame that $\gamma=A\sqrt{R}$ and $\beta=\sin\gamma/\gamma$, with the constant $A$ defined by Eq. (\ref{eq:critRadius}) \citep{1996ApJ...472..582M}.
%
%
\section{Time scales}
Since radiative cooling plays such an important role in our model (see the next section), we need to compare the cooling time scale to three other time scales related to the disk: the precession and growth time scales of the warp, and the shock passing time.
The latter is defined as the time the expanding wind blown bubble takes to travel to the outer disk radius $R_d$.

Adopting a cooling function of the form $\Lambda=\Lambda_1 T_s^{-1/2}$ and using the jump conditions for a strong shock, the cooling time of the shocked gas is given by \citep[cf.][]{1976A&A....50..145K, 1992ApJ...388..103K} $t_c = C v_s^3 / \rho_e$, with $v_s$ the shock speed, and $\rho_e$ the pre-shock environment density.
When assuming a fully ionized gas of cosmic abundances, the constant $C$ has a value of $6.0\times 10^{-35}$\,$\mathrm{g}\,\mathrm{cm}^{-6}\,\mathrm{s}^4$.

Analytical relations for the radius $R_s(t)$ and speed $v_s(t)$ of the outer shock as functions of time can be derived \citep[e.g.][]{1999isw..book.....L} and with these the shock passing time readily follows from setting $R_s(t_{sp})=R_d$ as
\begin{equation}
t_{sp} = \left(2/3\pi\rho_{e}\right)^{1/2} \dot{M}_w^{-1/2} v_w^{-1/2} R_d^2 \;.
\end{equation}
The precession time scale of the disk is given by \citep{1996ApJ...472..582M}
\begin{equation}
t_p=48\pi^2 c G^{1/2} M_*^{1/2} L_*^{-1} R_d^{3/2} \Sigma_d \; ,
\end{equation}
where we assumed Keplerian rotation.
The time scale for the initial growth of the warp is of the same order.
When we use Eq. (\ref{eq:critRadius}) for the critical radius as a typical disk radius, we find that the different time scales scale as
\begin{eqnarray}
t_p    & \propto & M_*^2 L_*^{-4} \dot{M}_{acc}^3 \eta^3 \Sigma_d \\
t_{sp} & \propto & \rho_e^{1/2} \dot{M}_w^{-1/2} v_w^{-1/2} M_*^2 L_*^{-4} \dot{M}_{acc}^4 \eta^4 \\
t_c    & \propto & \rho_e^{-5/2} \dot{M}_w^{3/2} v_w^{3/2} M_*^{-3} L_*^{6} \dot{M}_{acc}^{-6} \eta^{-6}
\end{eqnarray}

We are quite limited in our choice of $M_*$, $L_*$, $v_w$, and $\dot{M}_w$ since values for these parameters are strongly constrained by stellar evolution and wind models \citep[e.g.][]{1988A&A...207..123P, 1995A&A...299..755B} but since the dependence of $t_p$, $t_{sp}$, and $t_c$ on $\dot{M}_{acc}$ is so strong, a proper choice of this latter parameter leads to the desired proportion between the different time scales.

For our calculations we used $\dot{M}_w= 10^{-8}M_\odot\,\mathrm{yr}^{-1}$, $v_w=200\:\mathrm{km}\,\mathrm{s}^{-1}$, $M_*=0.6 M_\odot$, $L_*=10^4 L_\odot$, $\rho_e=10^{-15}\mathrm{g}\,\mathrm{cm}^{-3}$, $\dot{M}_{acc}= 10^{-7}M_\odot\,\mathrm{yr}^{-1}$, $\Sigma_d=1\mathrm{g}\,\mathrm{cm}^{-2}$, and $\eta=1$ resulting in $R_{crit}\simeq 2\mathrm{AU}$, $t_p\simeq 17\mathrm{yr}$, $t_{sp}\simeq 0.4\mathrm{yr}$, $t_c\simeq 10^{-8}\mathrm{yr}$, and density contrast $\chi\equiv\rho_d/\rho_e\simeq 300$ with $\rho_d$ the disk density.
So $t_c \ll t_{sp} \ll t_p$ showing that cooling will indeed be of importance and that we can safely ignore the disk's precession.
%
%
\section{Mechanism}
\begin{figure}
\centering
\includegraphics[height=17cm]{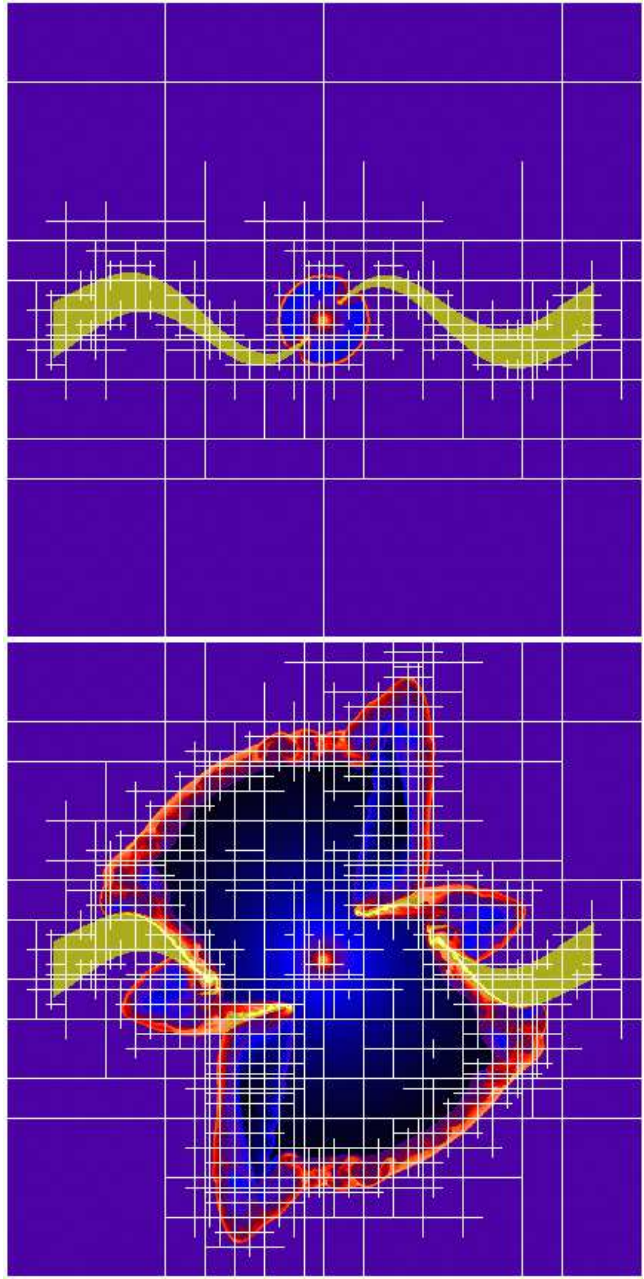}
\caption{2D example of a wind-disk interaction showing the evolving AMR grid structure superimposed on a plot of the logarithm of the density. Every square represents a grid of 8x8 cells. Five different levels of refinement are visible. The effective resolution of this simulation is 1024x1024 cells.}
\label{fig:grid}
\end{figure}

The mechanism behind the formation of the multipolar lobes seen in the simulations is as follows \citep[see also][]{2003A&A...405L..11I}.
As the central wind impinges on the inner rim of the disk, a three-dimensional bow shock develops around it.
This shock roughly has the shape of two `cones' connected by there tips at the disk's center.
The opening angle of the shock depends inversely on the Mach number of the wind.
Imagining a two-dimensional cut through the developing bow shock, one sees that one branch flies off into space creating a lobe jutting out from the nebula, whereas the other slams into the concave side of the disk, scooping up disk material and thereby producing a set of smaller, unstable lobes (see Fig.\ref{fig:grid}).
Due to the cooling of the gas, the swept up shell is highly compressed and therefore thin, and the ram pressure from the wind directly drives the shell outwards, which are necessary ingredients for the bow shock to produce the lobes.
Simulations of energy-driven flows never result in the pronounced point-symmetric lobes as seen in the calculations with cooling.
When the density of the disk is not too high, the wind breaks through the concave part of the disk, producing another pair of lobes.

Since the true bow shock is three-dimensional and the disk is warped, rotating the two-dimensional cut around the center of the disk shows that the concave side of the disk turns into the convex side and vice versa emphasizing the necessity of fully three-dimensional simulations of this interaction to truly understand the emerging point-symmetric structure.
%
%
\section{Numerical implementation}
We used the three-dimensional hydrocode {\em Flash} \citep{2000ApJS..131..273F} to model the interaction between a spherical wind and a warped disk.
This parallelized code implements block-structured AMR \citep{Berger1984, Berger1989} and a PPM type hydrosolver \citep{Woodward1984, Colella1984}.

Besides implementing the proper initial and boundary conditions, we also added our own cooling module in order to model radiative cooling using a cooling curve \citep{1972ARA&A..10..375D, 2002A&A...395L..13M}.
This curve gives the energy loss rate as a function of temperature for a low density gas in collisional ionization equilibrium.
The radiative losses were implemented through operator splitting and if the hydro timestep was larger than the cooling time, the former was subdivided into smaller steps when calculating the cooling.

To construct the warped disk, Eq.(\ref{eq:critRadius}) was combined with a constant `wedge angle' $\theta_d$ and a proper value for $A$, i.e. $R_d$ was taken to be a few times $R_{crit}$, see Eq. (\ref{eq:diskSurface}).
This disk was given a constant density which, through the density contrast $\chi$, resulted in a value for the environment density $n_e$.
The spherical wind was implemented as an inner boundary condition and given a $1/r^2$ density profile and a constant wind velocity $v_w$.
The pressure was calculated from an equation of state with a constant Poisson index $\gamma$.
%
%
\section{2D wind-disk simulations}
To check the implementation of the wind-disk interaction model into the AMR code described above, we repeated a number of the two-dimensional calculations previously done by \citet{2003A&A...405L..11I} (see Fig.\ref{fig:grid}).
Because he used a different hydrosolver (FCT/LCD) and the outcome of the calculations strongly depend on turbulent processes in the gas, the simulations did not agree in every single detail but the overall point-symmetric morphologies were retrieved.
All these simulations were run using a small value for the Poisson index ($\gamma=1.1$) resulting in `momentum driven' bubbles.

To see what happens when more realistic cooling is applied, we ran some simulations with the cooling curve module and a Poisson index $\gamma=5/3$.
This showed that, apart from the production of the by now familiar point-symmetric lobes, the outer shell of swept up gas is thinner and unstable and developed into a number of smaller lobes merging with one another as the shell expanded.
%
%
\begin{figure}
\centering
\includegraphics[height=17cm]{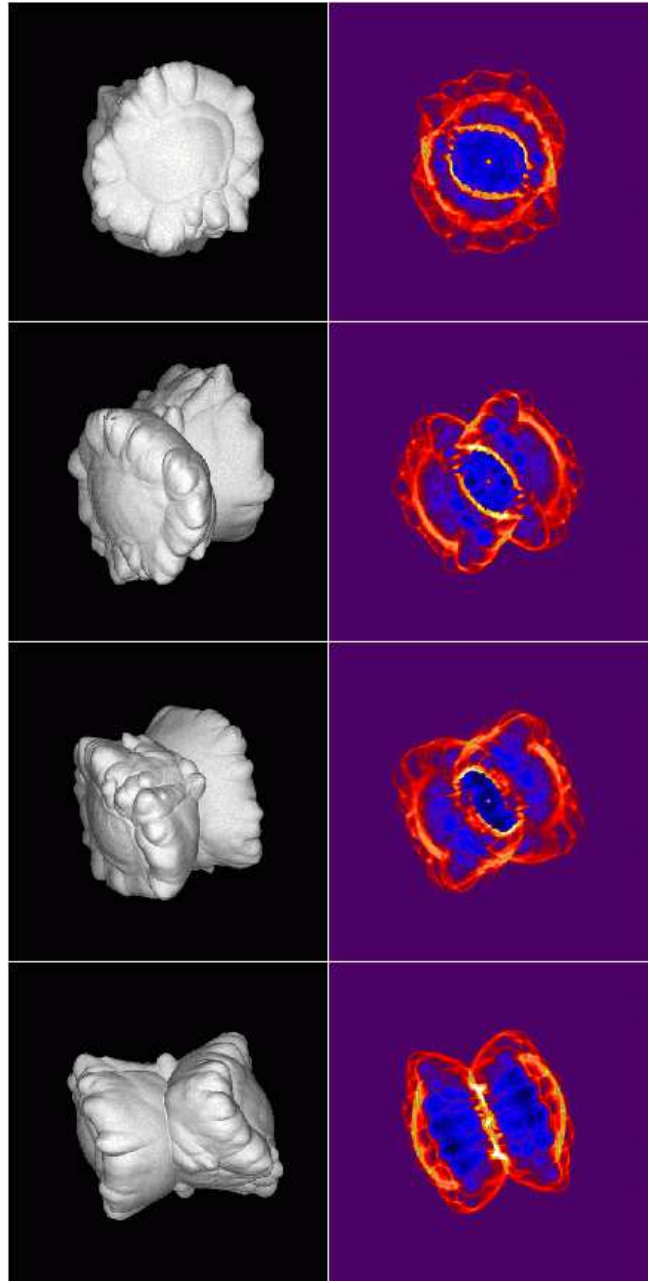}
\caption{Isosurfaces of the density at the end of the wind-disk simulation as seen from different angles ({\it left column}) and the corresponding synthesized $H\alpha$ images ({\it right column}).}
\label{fig:iso_proj}
\end{figure}
\section{3D AMR Simulations}
Following the two-dimensional trial calculations, we ran wind-disk simulations in three dimensions on a cartesian grid with an effective resolution of up to $512^3$ cells using five levels of refinement.
Since we found in our two-dimensional calculations that simulations with cooling applied through a cooling curve do not result in a qualitatively different morphological outcome compared to calculations with a low value for the Poisson index ($\gamma=1.1$), we opted not to use the cooling curve module for our three-dimensional simulations to save computational time.

We used the following parameters: $\gamma=1.1$, $n_e = 5\times 10^{8}\,\mathrm{cm}^{-3}$, $\chi=100$, $\theta_d=5^o$, and $v_w=200\:\mathrm{km}\,\mathrm{s}^{-1}$.
In Fig.\ref{fig:iso_proj} we present a visualization of the three-dimensional shape of the swept up shell through isosurfaces at different viewing angles.
Also shown are the corresponding synthesized $H\alpha$ images, derived by projecting the three-dimensional data cube onto the plane of the sky.
For this, we simply integrated the density squared along the line of sight and used this as a rough estimate for the emission.
%
%
\section{Conclusions}
Our computations show that the wind-disk interaction model in which the confining disk is warped results in a wide variety of point-symmetric shapes.
Nebulae that show 'punched holes', such as NGC~7027 \citep{2002A&A...384..603C}, are readily accommodated in this model.
Other candidates for our model are probably NGC~6537 and, to a lesser extent, NGC~7026, in which the inner disk is still visible, while the outer nebula shows clear point-symmetric structures.
Also, the shapes of the (proto-)PNe He~2-155 and M~4-18 (and maybe He~2-47) can be explained with our model.
As a further application, large-scale explosions in non-planar disks, such as might occur in active galaxies, are expected to show similar patterns, provided that the disk material can cool rapidly enough.
Movies of these simulations can be found at \url{http://www.strw.leidenuniv.nl/AstroHydro3D/}.
%
%
\subparagraph{Acknowledgements}
V.I. expresses his gratitude to Raghvendra Sahai and Hugo Schwarz for lively discussions that were the primary cause for taking up this subject.

The research of G.M. has been made possible by a fellowship of the Royal Netherlands Academy of Arts and Sciences.

The software used in this work was in part developed by the DOE-supported ASCI/Alliance Center for Astrophysical Thermonuclear Flashes at the University of Chicago.

Our work was sponsored by the National Computing Foundation (NCF) for the use of supercomputer facilities, with financial support from the Netherlands Organization for Scientific Research (NWO), under grant number $614.021.016$.
%
%
\bibliographystyle{aa}
\bibliography{rijkhorstAMR2003}
%
%
%
%
\end{document}